\documentclass[fleqn,usenatbib]{mnras}

\usepackage{newtxtext,newtxmath}

\usepackage[T1]{fontenc}

\DeclareRobustCommand{\VAN}[3]{#2}
\let\VANthebibliography\thebibliography
\def\thebibliography{\DeclareRobustCommand{\VAN}[3]{##3}\VANthebibliography}

\usepackage{graphicx}	
\usepackage{amsmath}	
\usepackage{amssymb}	
\usepackage{animate}
\usepackage{ulem}

\usepackage{bm}
\defcitealias{simard2011catalog}{S11}
\defcitealias{meert2015catalogue}{M15}
\defcitealias{dominguez2022sdss}{D22}

\title[The classical and pseudo-bulge galaxies]{Resolved properties of classical bulge and pseudo-bulge galaxies}

\author[Hu et al.]{
Jia Hu,$^{1,2}$\thanks{E-mail: hujia@nao.cas.cn}
Lan Wang,$^{1,2}$\thanks{E-mail: wanglan@bao.ac.cn}
Junqiang Ge,$^{3}$ 
Kai Zhu$^{4}$ and 
Guangquan Zeng$^{1,2}$
\\
$^{1}$Key Laboratory for Computational Astrophysics, National Astronomical Observatories, Chinese Academy of Sciences, Beijing 100101, China \\
$^{2}$School of Astronomy and Space Science, University of Chinese Academy of Sciences, Beijing 100049, China \\
$^{3}$National Astronomical Observatories, Chinese Academy of Sciences, Beijing 100101, China \\
$^{4}$Department of Astronomy, Tsinghua University, Beijing 100084, China\\
}

\date{Accepted XXX. Received YYY; in original form ZZZ}

\pubyear{2023}

\begin{document}
\label{firstpage}
\pagerange{\pageref{firstpage}--\pageref{lastpage}}
\maketitle

\begin{abstract}
We compare properties of classical and pseudo-bulges and properties of their hosting galaxies selected from the MaNGA survey. Bulge types are identified based on the S$\mathrm{\acute{e}}$rsic index n of bulge component and the position of bulges on the Kormendy diagram. For the 393 classical bulges and 422 pseudo-bulges selected and their hosting galaxies, we study their kinematic properties including a proxy for specific angular momentum and central velocity dispersion, their stellar population properties including stellar age, metallicity, and specific star formation rate, as well as HI fractions of the galaxies.  
Our results show that at given stellar mass, disc components of pseudo-bulge galaxies are younger, have more active star formation, rotate more, and may contain more HI content compared with those of classical bulge galaxies, and the differences are larger than those between bulges themselves. 
The correlations between bulge types and disc properties indicate that different types of bulges are shaped by different processes that may regulate both growth of central components and evolution of outer discs in galaxies.
In addition, we propose a stellar mass dependent divider of central velocity dispersion to separate galaxies with classical bulges from those with pseudo-bulges in galaxy mass range of $10.4<\mathrm{log}(M_*/\rm M_\odot)<11.4$: $\mathrm{log}(\sigma_0) = 0.23 \times \mathrm{log}(M_*/\rm M_\odot)-0.46$. Galaxies with larger/smaller $\sigma_0$ can be classified as hosts of classical/pseudo-bulges.
\end{abstract}



\section{Introduction}
\label{sec:Intro}

A galactic bulge is a central component that exists commonly in a spiral galaxy, and has more concentrated light and stars than the disc component that extends to a larger radius \citep{1961hag..book.....S}. 
Observations indicate that there are two main categories of bulges: classical bulges and pseudo-bulges. 
Typical classical bulges are considered to be analogous to elliptical galaxies, being random motion dominated and following the fundamental plane \citep{2016ASSL..418...41F}. 
On the other hand, pseudo-bulges resemble spiral galaxies that are rotation dominated, and typically have substructures such as nuclear bars, spiral arms, and inner rings \citep{1997AJ....114.2366C,2002AJ....124...65E,2005A&A429.141K,2011ApJ...733L..47F}. 
Compared with classical bulges, pseudo-bulges are in general bluer in color, and have more active star formation activities \citep{2007ApJ...664..640D, 2012ApJ...754...67F, 2020ApJ...899...89S}. 

Theoretically, classical bulges are thought to form through violent processes such as clumpy collapse \citep{1999ApJ...514...77N,2008ApJ...688...67E,2012MNRAS.422.1902I} and galaxy merger \citep{2001A&A...367..428A, 2005A&A...430..115H}, while pseudo-bulges are considered to be shaped slowly by secular evolution \citep{1996ApJ...457L..73C,2004ARA&A..42..603K,2005MNRAS.358.1477A}, such as mild gas inflow and stellar inward migration induced by bar or spiral arm \citep{2010MNRAS.407L..41S,2012MNRAS.420..913B,2013ApJ...772...36G,2014MNRAS.439..623G,2015A&A...578A..58H,2020IAUS..353..155C,2022MNRAS.512.2537G}. 
Nevertheless, detailed studies of bulge formation indicate a more complicated picture, both from simulation / semi-empirical models 
\citep{2010ApJ...715..202H,2013ApJ...772...36G,2015A&A...579L...2Q,2018MNRAS.473.2521S,2016MNRAS.462L..41F}, and from observation \citep{2008ApJ...687...59G,2013A&A...552A..67E,2016ASSL..418...77L}.

If different properties of bulges are indeed determined by various formation processes, for example violent mergers or long-term effects, galaxies hosting them should have different properties. 
Efforts have been made in this direction for galaxies in the local Universe.
For example, \citet{2019MNRAS.484.3865W} found that pseudo-bulge galaxies and low mass classical bulge galaxies have more close neighbour galaxies when compared to control galaxies, implying strong connections between galactic bulges and galaxy–galaxy interactions.
\citet{2022A&A...661A..98Y} found that galaxies with stronger spiral arms tend to have an increasing fraction of pseudo-bulges, which suggests that spiral arms play a role in the secular buildup of pseudo-bulges.

With the help of integral-field spectroscopic surveys such as CALIFA \citep{2012A&A...538A...8S} and MaNGA \citep{2015ApJ...798....7B}, spatially resolved properties for galaxies with bulges have been studied. 
For example, \citet{2018MNRAS.481.5580F} and \citet{2022MNRAS.514.6141J} investigate age and metallicity of stellar populations in bulge and disc components of S0 galaxies in MaNGA. 
\citet{2022MNRAS.514.6120J} studied the age of two-component galaxies in the MaNGA survey. In these studies, bulges are found to have similar or larger age and metallicity than discs. 
\citet{2023A&A...669A..70B} studied spatially resolved star formation history and properties of stellar population including colour, age and metallicity, of 135 late-type galaxies in CALIFA. 
They claimed that pseudo and classical bulges form a continuous sequence and are not entirely different populations.

In this work, in order to investigate further the formation mechanism of bulges, we study the difference between properties of classical and pseudo-bulges, as well as the difference between properties of their host galaxies. Based on the large sample of MaNGA survey ($\sim$10,000 galaxies), we are able to select galaxies with different types of galactic bulges, which include enough numbers to be compared statistically. Apart from the general properties of bulges and galaxies, spatially resolved kinematic and stellar population properties are studied, and radial profiles of these properties are presented, to reveal the differences in more detail.

This paper is organized as follows. Section 2 introduces the MaNGA data and how we select galaxies with different types of bulges from a decomposition catalog based on it. Section 3 studies and compares the kinematic properties of bulges and galaxies, focusing on the index of $\lambda$ that indicates a proxy for the specific angular momentum of a system, as well as central velocity dispersion $\sigma_0$. Section 4 studies properties of stellar population including age, metallicity and specific star formation rate (sSFR), for both bulge component and the whole galaxy. Conclusions and discussions are presented in Section 5.

\section{Sample selection}
\label{sec: Sample selection}

For MaNGA galaxies, based on the decomposition catalogue of \citet{2022MNRAS.509.4024D}, we select galaxies with classical bulge and pseudo-bulge by criteria of the bulge S$\mathrm{\acute{e}}$rsic index combined with position of bulges on the Kormendy diagram \citep{2009MNRAS.393.1531G}. 

\subsection{MaNGA galaxies}
\label{sec: MaNGA and MaNGA galaxies}

The Mapping Nearby Galaxies at Apache Point Observatory \citep[MaNGA,][]{2015ApJ...798....7B} is an integral field spectroscopic survey based on the 2.5-meter telescope at the Apache Point Observatory \citep{2006AJ....131.2332G}, which is one of the three main projects in the fourth-generation Sloan Digital Sky Survey \citep[SDSS,][]{2017AJ....154...28B}. Using fiber bundles, the MaNGA survey provided spatially resolved spectra for $\sim$10,000 galaxies with a wide distribution of stellar masses and colour, covering redshifts in the range of 0.01 < $z$ < 0.15 \citep{2016AJ....152..197Y,2017AJ....154...86W}.

In this work, we make use of the two-dimensional stellar kinematics (i.e. the maps of stellar velocity and stellar velocity dispersion) provided by the Data Analysis Pipeline \citep[DAP, ][]{2019AJ....158..231W}, which are derived using the \textsc{ppxf} \citep{2004PASP..116..138C,2017MNRAS.466..798C,2023MNRAS.526.3273C} software combined with the \textsc{mastar} stellar library \citep{2019ApJ...883..175Y}. To ensure that the measured stellar velocity dispersion are reliable, the IFU spectra had been Voronoi binned \citep{2003MNRAS.342..345C} to reach $\mathrm{S/N} \sim 10$ before spectra fitting \citep{2017AJ....154...86W}. Specifically, the stellar velocity dispersion used in this work has been corrected for the instrumental velocity dispersion.

In the final release of MaNGA \citep[SDSS DR17, ][]{2022ApJS..259...35A}, DAP includes 10782 sources, among which 637 sources are Coma, IC342, M31, globular clusters and other non-galactic sources, and 135 sources have been observed repeatedly. 
The remaining 10,010 galaxies are defined as unique (with the unique MaNGA-ID). We have checked that there are still 9 pairs of identical sources with different MaNGA-IDs and we keep only one of them. In the end, 10,001 galaxies are unique in MaNGA. Of the 10,001 galaxies selected above, 5 failed to be Voronoi binned, and our following analysis is based on the rest 9996 galaxies.

\subsection{Decomposition catalogue and two-component galaxies}
\label{sec: Decomposition catalogue and two-component galaxies}

S$\mathrm{\acute{e}}$rsic index criterion \citep{2004ARA&A..42..603K,2008AJ....136..773F} is commonly used to select classical bulge and pseudo-bulge, and the result relies on the robustness of photometric decomposition of bulge and disc components. 
For MaNGA galaxies, several decomposition catalogues are available, as presented by \citet{2011ApJS..196...11S} and \citet{2015MNRAS.446.3943M} which give results for larger samples of SDSS galaxies, and by \citet{2022MNRAS.509.4024D} for MaNGA galaxies in particular.

We have compared the decomposition results from these three catalogues in detail, and choose the one provided by \citet{2022MNRAS.509.4024D}, the MaNGA PyMorph photometric Value Added Catalogue (hereafter MPP-VAC-DR17), to identify two-component galaxies and further different bulge types.
\citet{2011ApJS..196...11S} overestimates sky background of galaxies with large size \citep{2015MNRAS.446.3943M}, which affects their decomposition results and bulge type identification.
\citet{2015MNRAS.446.3943M} and \citet{2022MNRAS.509.4024D} both use analysis pipeline PYMORPH \citep{2010MNRAS.409.1379V} and obtain similar results for most galaxies. We choose the latter for several reasons. 
First, the catalogue of \citet{2015MNRAS.446.3943M} only includes $\sim$85\% of MaNGA galaxies, while MPP-VAC-DR17 includes 99.8\%. 
Second, MPP-VAC-DR17 performed extra refit of galaxies that were not fitted well. 
Besides, we have inspected the images of MaNGA galaxies and find that galaxies defined as two-component galaxies in MPP-VAC-DR17 indeed exhibit bright centrals. 
Therefore MPP-VAC-DR17 is believed to provide more robust results for selecting two-component galaxies.
Out of the 9,996 galaxies selected in \S \ref{sec: MaNGA and MaNGA galaxies}, 9975 galaxies are included in MPP-VAC-DR17.

Following the method of \citet{2015MNRAS.446.3943M} and \citet{2019MNRAS.483.2057F}, the MPP-VAC-DR17 provided by \citet{2022MNRAS.509.4024D} performed two-dimensional fits for each MaNGA galaxy image in the g-, r-, and i-bands respectively using two models: a single S$\mathrm{\acute{e}}$rsic profile \citep[]{1963BAAA....6...41S}, and a two-component model consisting of a S$\mathrm{\acute{e}}$rsic profile for the bulge and an exponential profile for the disc.
The S$\mathrm{\acute{e}}$rsic profile has a form of $I(R)\propto$ $\rm exp$$[-b_n(R/R_e)^{\frac{1}{n}}]$, with parameters of effective radius $R_e$, S$\mathrm{\acute{e}}$rsic index $n$, and $b_n$, a dimensionless constant that depends on $n$ \citep{1989woga.conf..208C,2005PASA...22..118G}. The detailed equations used can be found in equation (4) of \citet{2015MNRAS.446.3943M}.
In the catalogue, a visual-inspection-based flagging system gives a parameter FLAG\_FIT to evaluate which model is to be preferred for scientific analyses. Galaxies that are identified as single-component galaxies have FLAG\_FIT=1 and galaxies with disc+bulge components have FLAG\_FIT=2. 

There are 3170 galaxies with FLAG\_FIT=2 for r-band image in the MPP-VAC-DR17. 
We have checked further the decomposition results for these galaxies in detail, and exclude some of the galaxies that may not be fitted reasonably, in order to get a more robust sample. 
Firstly, 51 galaxies have larger bulge components than disc components, with larger effective radius for bulge than for disc. 
Secondly, 682 galaxies have bulges with effective radius smaller than the half-width at half-maximum (HWHM) of point spread function (PSF) of SDSS images (0.75$\arcsec$), which can not be identified clearly \citep{2009MNRAS.393.1531G}. 
Finally, 2437 bulge+disk galaxies are selected to be analysed in this work.

In addition, we also select elliptical galaxies in the MPP-VAC-DR17, to set Kormendy relation criterion for identifying bulge types in the coming subsection. 
For the galaxies with FLAG\_FIT=1 in r-band that are considered to be signal-component ones, 1698 galaxies have large S$\mathrm{\acute{e}}$rsic index ($3<n<7.95$) and are selected as elliptical ones. Besides,
\citet {2022MNRAS.509.4024D} also provides galaxy morphology results of machine learning (the MaNGA Deep Learning Morphological Value Added Catalogue). According to this catalogue, we select elliptical galaxies to be early-type with $\rm T-Type<0$ and have a low probability of being S0 galaxy($P_{\rm S0}<0.5$), to get a further constrain. 
All combined, 1293 typical elliptical galaxies are selected.

\subsection{Selection of classical and pseudo-bulge galaxies}
\label{sec: Selection of classical and pseudo-bulge galaxies}

Based on the two-component galaxy sample selected as described above, we further select galaxies with classical bulges and pseudo-bulges. S$\mathrm{\acute{e}}$rsic index of bulge component $n_b$ is the most commonly used criterion to select bulge types, since studies of nearby galaxies with robust bulge classification show that most classical bulges have $n_b>2$ and most pseudo-bulges have $n_b<2$ \citep[e.g.][]{2004ARA&A..42..603K,2016ASSL..418...41F}. 
Following \citet{2019MNRAS.484.3865W}, to minimize the effect of uncertainty in fitting bulge S$\mathrm{\acute{e}}$rsic index, we select 1360 galaxies with $0<n_b<2$ to be pseudo-bulge host candidates and 524 galaxies having $3<n_b<7.95$ to be classical bulge host candidates. 

In addition, positions of bulges on the Kormendy diagram \citep{1977ApJ...218..333K} are also often used to select more reliable and typical bulges of different types \citep{2015MNRAS.450..873V,kormendy2016elliptical,2017MNRAS.472L..89M}. Kormendy relation is a tight correlation between effective radius $R_e$ and average surface brightness within the effective radius $\langle \mu_e \rangle$ for elliptical galaxies, as a projection of fundamental plane. Classical bulges are found to follow the Koremendy relation, while pseudo-bulges normally deviate from this relation \citep{2009MNRAS.393.1531G}.
We have checked that if samples are selected just using the Sersic $n_b$ criterion, the differences between bulge samples shown in the following are a bit smaller, but the results of the paper remain unchanged.

\begin{figure}
	\includegraphics[width=\columnwidth]{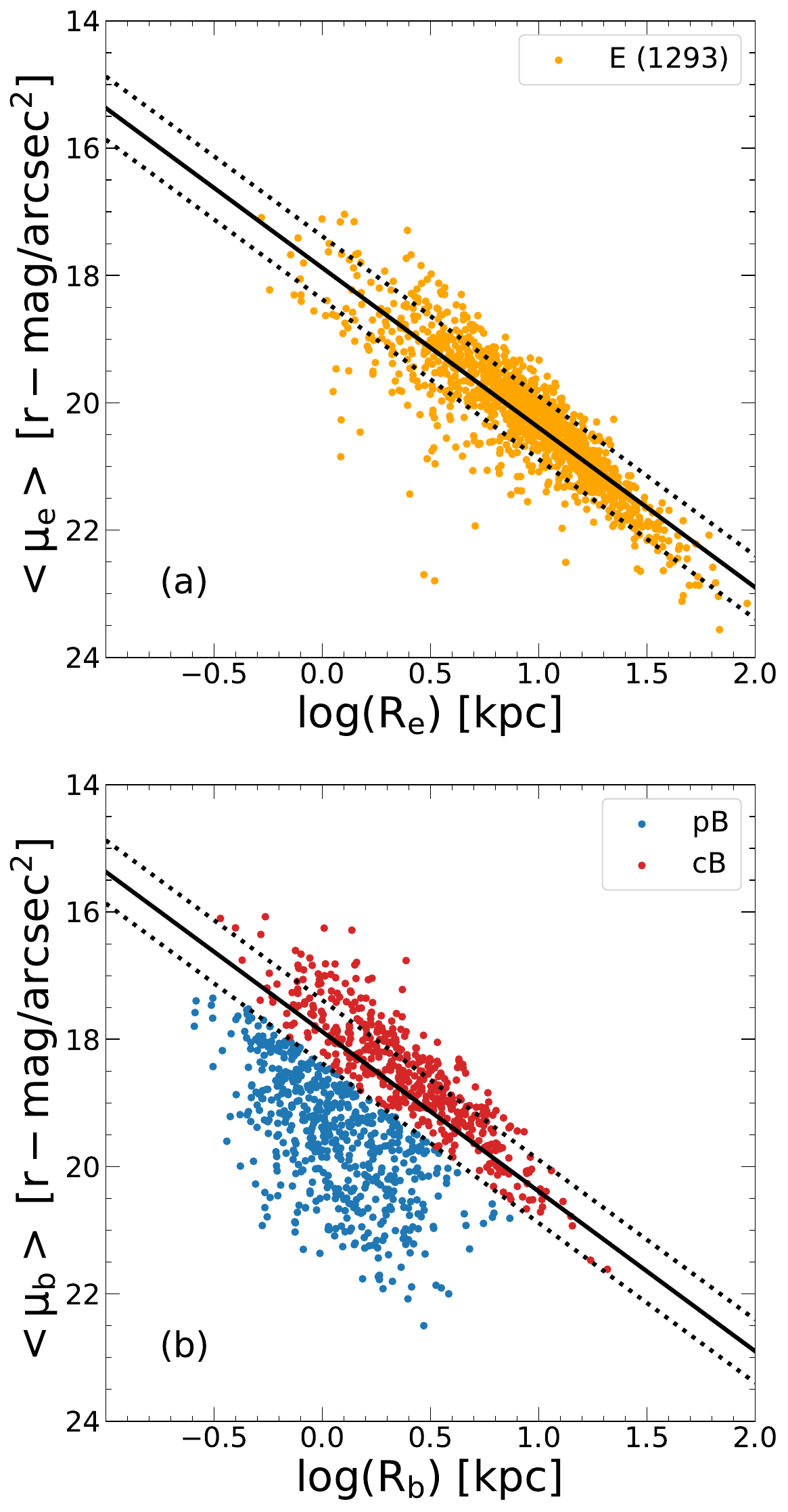}
    \caption{Panel (a): $r-$band Kormendy Relation of our selected elliptical galaxies (E). The orange dots represent individual elliptical galaxies. The black solid line is the best linear fitting of these dots. The two black dotted lines show the $\pm 1\sigma$ scatter. Panel (b): In the Kormendy diagram, the position of classical bulges (cB: red dots) and pseudo-bulges (pB: blue dots) is constrained by both $n_b$ and Kormendy relation. }
    \label{fig:kormendy relation}
\end{figure}

Panel (a) of Fig.~\ref{fig:kormendy relation} shows the Kormendy relation of elliptical galaxies as described in \S \ref{sec: Decomposition catalogue and two-component galaxies}. The black solid line is the best linear fitting result:
\begin{equation}
    \langle \mu_e \rangle  = 2.51 \times \mathrm{log}(R_e/\mathrm{kpc})+17.88.
    \label{eq:eq3}
\end{equation}
The two black dotted lines show the $ 1\sigma$ scatter ($ \sigma$=0.50). 
For galaxies with bulges, we get the bulge effective radius $R_b$ from MPP-VAC-DR17, which is from the two-component fitting result as described in section 2.2, and calculate the average surface brightness within bulge effective radius $\langle \mu_b \rangle$, to plot positions of bulges on the Kormendy diagram. To get a typical sample of pseudo-bulges, in the Kormendy diagram, we require the selected bulges to lie below the lower dotted line in panel (a) of Fig.~\ref{fig:kormendy relation}, which reduces the number of pseudo-bulge galaxies selected by  S$\mathrm{\acute{e}}$rsic index by a large fraction, leaving a sample of 567 galaxies. 
The pseudo-bulges constrained by both S$\mathrm{\acute{e}}$rsic index and Kormendy criterion are plotted in panel (b) of Fig.~\ref{fig:kormendy relation} by blue dots. 
For classical bulges, we require them to lie above the lower dotted line in the relation set by ellipticals. 
This results in a sample of 459 classical bulges, as indicated by red dots in panel (b) of Fig.~\ref{fig:kormendy relation}. 

We obtain the spatially resolved properties of stellar components for our selected galaxies, using the PPXF code for full spectrum fitting, Vazdekis/MILES model \citep{2010MNRAS.404.1639V} to generate simple stellar population templates by assuming the BaSTI isochrone and Salpeter IMF \citep{ge2018, 2019MNRAS.485.1675G, 2021MNRAS.507.2488G}.
Stellar mass, star formation rate, as well as stellar age and metallicity are calculated for each pixel of a galaxy. 
For our classical (pseudo) bulge galaxy samples, we exclude 3 (3) galaxies that fail in the stellar population analysis, 2 (0) galaxies that are off-center, and 19 (14) galaxies that are affected by signals from foreground stars or neighbor galaxies within an ellipse of $R_e$. 
In addition, as this study primarily focuses on the bulge components, we exclude 42 classical and 128 pseudo-bulges galaxies whose effective radii of bulges are smaller than the HWHM of PSF in MaNGA (with a mean value of 1.25$\arcsec$). 
Finally, 393 classical bulge galaxies and 422 pseudo-bulge galaxies are used in the following for further analysis.

\begin{figure}
	\includegraphics[width=\columnwidth]{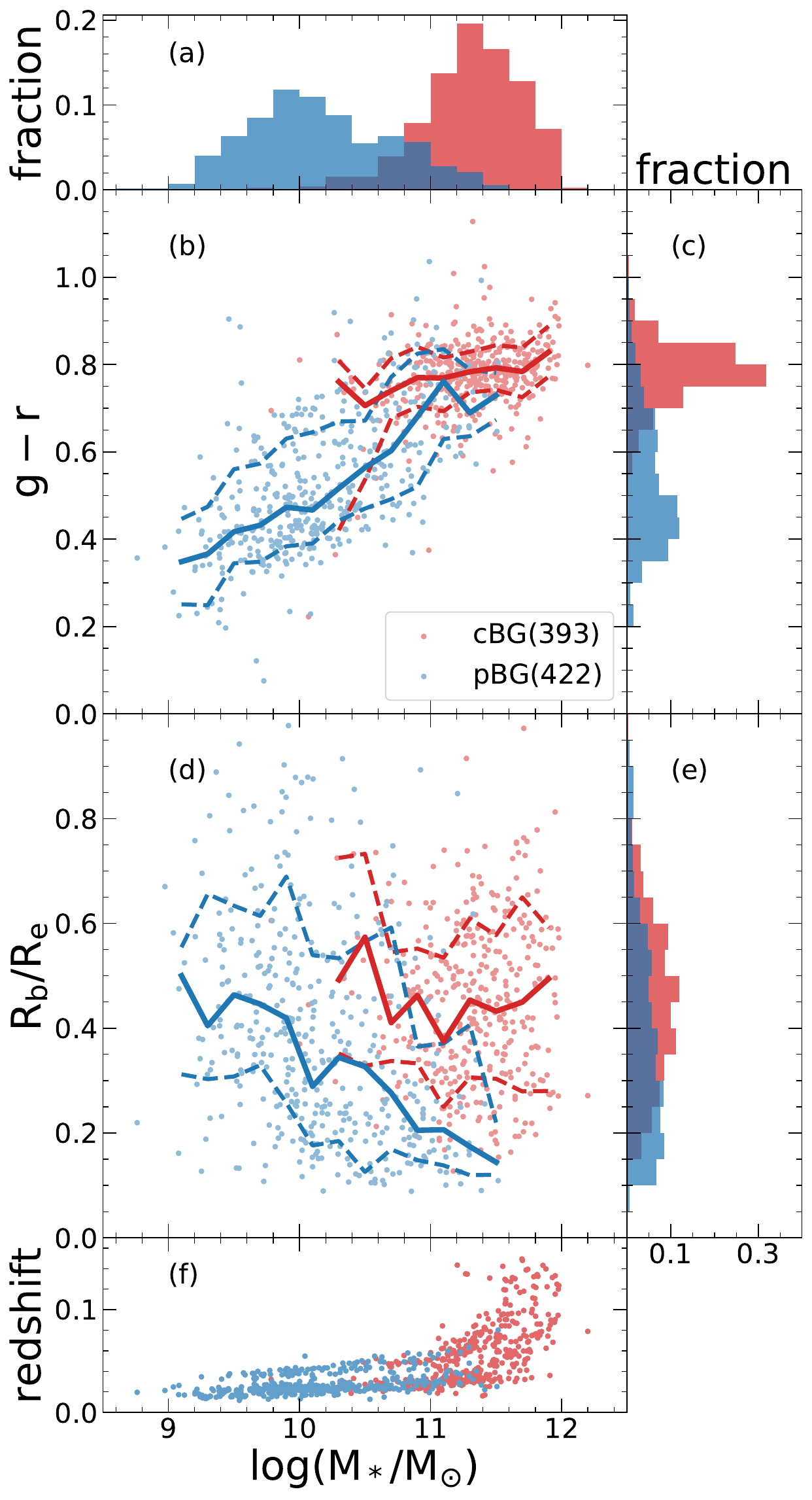}
    \caption{General properties of our selected classical bulge galaxies (cBG: red symbols) and pseudo-bulge galaxies (pBG: blue symbols). Panels (b),(d) and (f) present g - r colour, bulge to total size ratio $R_b/R_e$, and redshift, as a function of galaxy total stellar mass. Each dot indicates an individual galaxy. Solid lines show the median relation and dashed lines enclose 68\% scatter.
    Panels (a), (c), and (e) present the distributions of galaxy total stellar mass, colour, and bulge to total size ratio respectively for the two samples.}
    \label{fig: g-r-m-z}
\end{figure}

In Fig.~\ref{fig: g-r-m-z}, we compare the general properties of galaxies with different types of bulges, including g - r colour, bulge to total size ratio ($R_b/R_e$) and redshift, as a function of galaxy total stellar mass. 
The g-, r-band apparent magnitude, and redshift are provided by NASA-Sloan Atlas (NSA) catalogue \footnote{\url{https://www.sdss.org/dr13/manga/manga-target-selection/nsa/}}. 
The effective radii of bulges and whole galaxies are from the MPP-VAC-DR17. 
For galaxy stellar mass, following \citet{2021MNRAS.507.2488G}, twice of $M_{R_e}$ is used to represent galaxy total stellar mass $M_*$. 
$M_{R_e}$ is calculated as the sum of each pixel stellar mass within an ellipse with $R_e$ and axis ratio of the whole galaxy.

Fig.~\ref{fig: g-r-m-z} displays that the selected classical bulge galaxies are in general more massive and redder than pseudo-bulge galaxies, consistent with previous studies\citep{2019MNRAS.484.3865W,2020ApJ...899...89S,2010ApJ...716..942F,2022MNRAS.515.1175H}. 
Panel (b) indicates an increase in colour with increasing stellar mass for pseudo-bulge galaxies, while classical bulge galaxies are mostly red with g - r$\sim$0.8. 
Panel (d) reveals that pseudo-bulge galaxies tend to have smaller relative sizes of bulges at larger stellar masses, while the relative bulge size of classical bulge galaxies does not vary much with mass, and is larger than that of pseudo-bulge galaxies of similar stellar mass, consistent with the previous study of \citet{2023A&A...680A..49Q}.  
Panel (f) shows that galaxies with classical bulges are in general at higher redshift, consistent with the larger mass they have.

\section{Kinematic properties}
\label{sec:kinematic}
 
In this section, we compare the general and spatially resolved kinematic properties of galaxies with different types of bulges within various stellar mass bins. 
The properties investigated include $\lambda_{R}$ that indicates proxy for specific angular momentum of systems, and central velocity dispersion ${\sigma}_{0}$. 
While classical bulge galaxies are redder than pseudo ones for given stellar mass, as shown in panel (b) of ~Fig. \ref{fig: g-r-m-z}, the difference that appears, if any, may be partially due to the different colours for these two samples. 
We have checked that when fixing both the total stellar mass and g - r colour, all the results shown below remain similar. 

\subsection{\texorpdfstring{$\lambda_{R}-\epsilon$}{Lg} relation}
\label{sec: rotational properties}

While pseudo-bulges are considered to be more rotationally supported than classical bulges \citep{2012ApJ...754...67F}, we first investigate in detail the kinematic rotational properties of classical and pseudo-bulges and their hosting galaxies. 

For IFU data as provided in MaNGA, $\lambda_{R}$ is widely used as a proxy for the specific stellar angular momentum of galaxy \citep{2007MNRAS.379..401E,2018MNRAS.477.4711G,2023MNRAS.522.6326Z,2024MNRAS.527..706Z}, with larger $\lambda_{R}$ reflecting faster rotation. 
For most disk galaxies, $\lambda_{R}$ increases with radius $R$ all the way to $R=R_e$ before getting flat \citep{2011MNRAS.414..888E, 2020MNRAS.495.4638O}. 
Therefore $\lambda_{R_e}$ can be used to represent global rotation of galaxies \citep{2007MNRAS.379..401E}. In practice, to take into account the fact that line-of-sight velocity is not corrected for the inclination angle for each galaxy, the position of a galaxy on the $\lambda_{R_e}-\epsilon$ diagram is normally adopted to reflect its global rotation \citep{2007MNRAS.379..401E}, where ellipticity $\epsilon$ is calculated based on axial ratio $b/a$ ($\epsilon=1-b/a$). 

Following \citet{2007MNRAS.379..401E}, we calculate $\lambda_{R}$ as:
\begin{equation}
    \lambda_{R} \equiv \frac{\langle R|V| \rangle}{\langle R\sqrt{V^2+\sigma^2} \rangle}= \frac{\sum_{i=1}^N F_i R_i |V_i|}{\sum_{i=1}^N F_i R_i \sqrt{V^2_i+\sigma^2_i}}
    \label{eq:lambda}
\end{equation}
which sums over $N$ pixels within an ellipse with radius $R$ and the axial ratio $b/a$. $F_i$, $R_i$, $V_i$, and $\sigma_i$ represent the mean flux, radius, line-of-sight velocity, and velocity dispersion of stars on the $ith$ pixel, respectively. 

\begin{figure*}
	\centering
	\includegraphics[width=2.0\columnwidth]{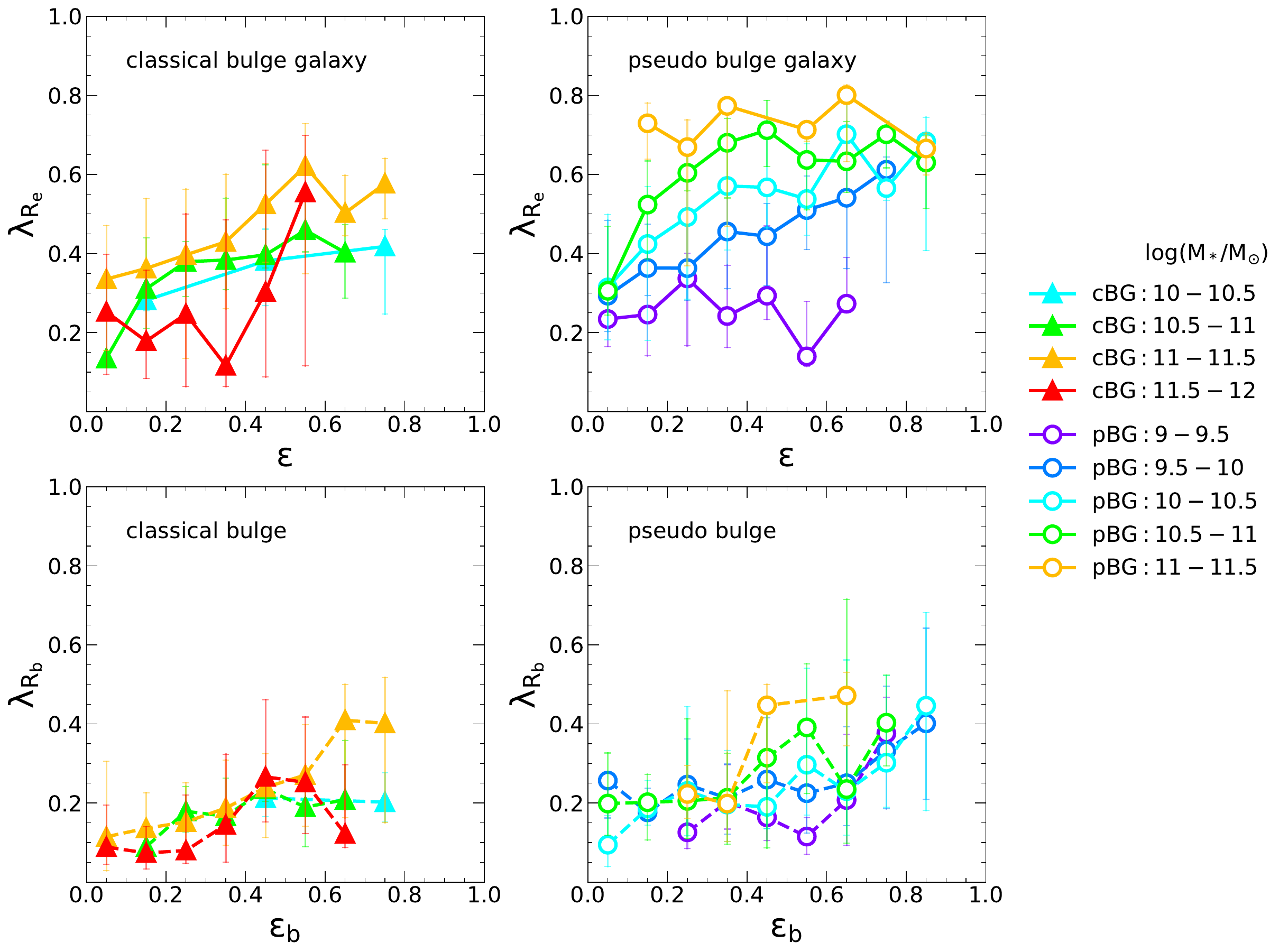}
	\caption{The $\lambda_{R_e}-\epsilon$ relation of the whole galaxies (top panels) and the $\lambda_{R_b}-\epsilon_b$ relation for the bulge components (bottom panels), for classical (left panels) and pseudo-bulge galaxies (right panels). Results are shown in galaxy stellar mass intervals of 0.5 in logarithmic space as indicated, and each symbol includes 3 galaxies at least. For each subsample, the median $\lambda_R$ is indicated by triangle or circular symbols, and the error bars show the $1\sigma$ scatter.}
    \label{fig: lambda-e in different mass}
\end{figure*}

In the top two panels of Fig.~\ref{fig: lambda-e in different mass}, we show the $\lambda_{R_e}-\epsilon$ relation for classical (top left) and pseudo-bulge (top right) galaxies in different galaxy stellar mass intervals.
In general, for galaxies with a given type of bulges, more massive galaxies have larger $\lambda_{R_e}$, indicating a faster rotation than lower mass galaxies, consistent with the results for late-type galaxies in CALIFA \citep{2015IAUS..311...78F}, SAMI \citep{2021MNRAS.508.2307V} and MaNGA \citep{2020MNRAS.495.1958W,2024MNRAS.527..706Z}.
When comparing classical bulge galaxies with pseudo-bulge galaxies, the latter rotates more for given stellar mass. 
Note that classical bulge galaxies in the most massive bin (red symbols in the top left panel) have less rotation than the galaxies within the second massive bin (orange symbols in the top left panel). 
We have checked in detail that this is partially attributed to a higher fraction (75.6\%) of galaxies without spiral arms or rings in the most massive bin than in the other stellar mass bins ($\sim$ 50\%). These galaxies are normally slow-rotating systems compared to galaxies with spiral arms or rings. 
In addition, in the most massive bin, more galaxies without spiral arms or rings are rotating extremely slow with $\lambda_{R_e} \sim 0.1$, with little velocity and large velocity dispersion both at small and large radii.
For galaxies with spiral arms or rings, more massive galaxies always rotate more for a given bulge type, and classical bulge galaxies rotate less than pseudo bulge ones at given stellar mass.

The bottom panels of Fig.~\ref{fig: lambda-e in different mass} show the $\lambda_{R_b}-\epsilon_b$ relation, for the bulge components of our selected samples. 
The ellipticity $\epsilon_b$ is calculated based on axial ratio $b/a$ of bulge component from MPP-VAC-DR17. 
As expected, bulges have slower rotation than the galaxies in general.
Unlike as seen in the top panels, bulges of each type show similar rotation state, almost independent of galaxy stellar mass. 
Comparing the bulges of different types, the results are quite similar in all stellar mass bins, with in general only a bit larger rotation for pseudo-bulges. 
Fig.~\ref{fig: lambda-e in different mass} shows clearly that different types of bulges have small differences in rotation, while their hosting galaxies as a whole have large differences. 
This could be partially due to the fact that classical bulges are relatively larger in size than pseudo-bulges as seen in panel (d) of Fig. ~\ref{fig: g-r-m-z}, and contribute more to the total rotation of galaxies. 
Apart from this, disc component of galaxies with different bulges may also rotate differently.  

\begin{figure*}
	\centering
	\includegraphics[width=2.0\columnwidth]{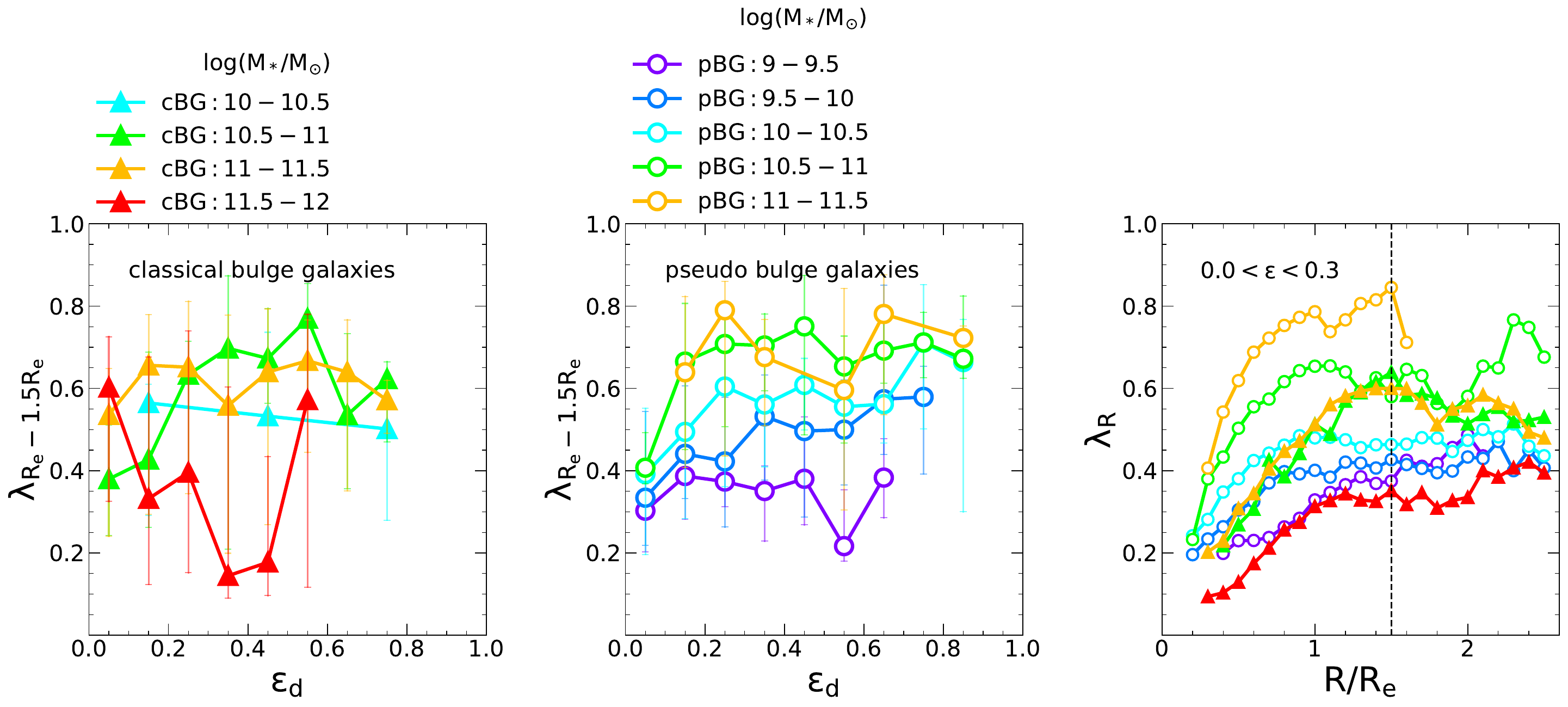}
	\caption{Left (middle) panel: $\lambda_R-\epsilon_d$ relation of the disc component for classical (pseudo) bulge galaxies in different stellar mass bins. Symbols indicate the median relation and error bars give the $1\sigma$ scatter. Right panel: For galaxies with $0.0<\epsilon<0.3$, the median $\lambda_R$ radial profile of classical and pseudo-bulge galaxies indicated by various symbols. In the right panel, symbols in radial intervals of more than 5 galaxies are shown, and the cyan triangles include too few galaxies in all radius bins to be shown. The vertical black dotted line marks the position of $1.5*R_e$.}
    \label{fig: lambda of disk different mass}
\end{figure*}

We check the rotation of disc components ($\lambda_R-\epsilon_d$ relation) of galaxies hosting classical bulges and pseudo-bulges, in the left and middle panel of Fig.~\ref{fig: lambda of disk different mass} respectively. 
To minimize the contribution of bulge components, we calculate $\lambda_R$ within an elliptical ring with radius from $R_e$ to $1.5R_e$, to quantify the rotation of disc components. 
Ellipticity of the disc component $\epsilon_d$ is from MPP-VAC-DR17.
At overlapped stellar mass intervals, pseudo-bulge galaxies have in general faster rotating discs than classical bulge galaxies, but the difference is smaller than that of galaxies as a whole, as presented in the upper panels of Fig.~\ref{fig: lambda-e in different mass}. Therefore the latter is largely affected by the larger relative size of classical bulges than pseudo-bugles, with also a contribution from slower disc rotation.

In the right panel of Fig.~\ref{fig: lambda of disk different mass}, we present the radial profiles of $\lambda_R$ for classical and pseudo-bulge galaxies in different stellar mass bins, to have a more detailed look at the rotation state of galaxies at various radii. 
While $\lambda_R$ depends on $\epsilon$, we show results for galaxies within a narrow range of $0.0<\epsilon<0.3$. 
We also check $\lambda_R$ profile of galaxies within other $\epsilon$ ranges and find similar results. 
Affected by the PSF of spectrum in MaNGA, innermost points of galaxies are chosen to be greater than their HWHM. 
For MaNGA galaxies, about 65\% are covered by the fibers to the radius of $1.5R_e$, with the rest going to $2.5R_e$. 
Therefore in this panel, more galaxies are included within $1.5R_e$, which is indicated by the vertical dotted line. 
We have checked that the results within $1.5R_e$ are similar if only include galaxies that have data extending to $2.5R_e$.

For galaxies more massive than $10^{10.5}\rm M_\odot$, those with pseudo-bulges rotate more than the ones with classical bulges at radius smaller than $1.5R_e$. 
The difference between two subsamples is largest at around $0.8R_e$. While the bulge size is mostly around $0.3R_e$ - $0.6R_e$ for classical bulges and $0.1R_e$ - $0.5R_e$ for pseudo-bulges as presented in Fig.~\ref{fig: g-r-m-z}, the right panel of Fig.~\ref{fig: lambda of disk different mass} shows more clearly that at stellar mass greater than $10^{10.5}\rm M_{\odot}$, galaxies that host pseudo-bulges have discs that are rotating more than the ones hosting classical bulges.

\begin{figure*}
	\centering
	\includegraphics[width=2.0\columnwidth]{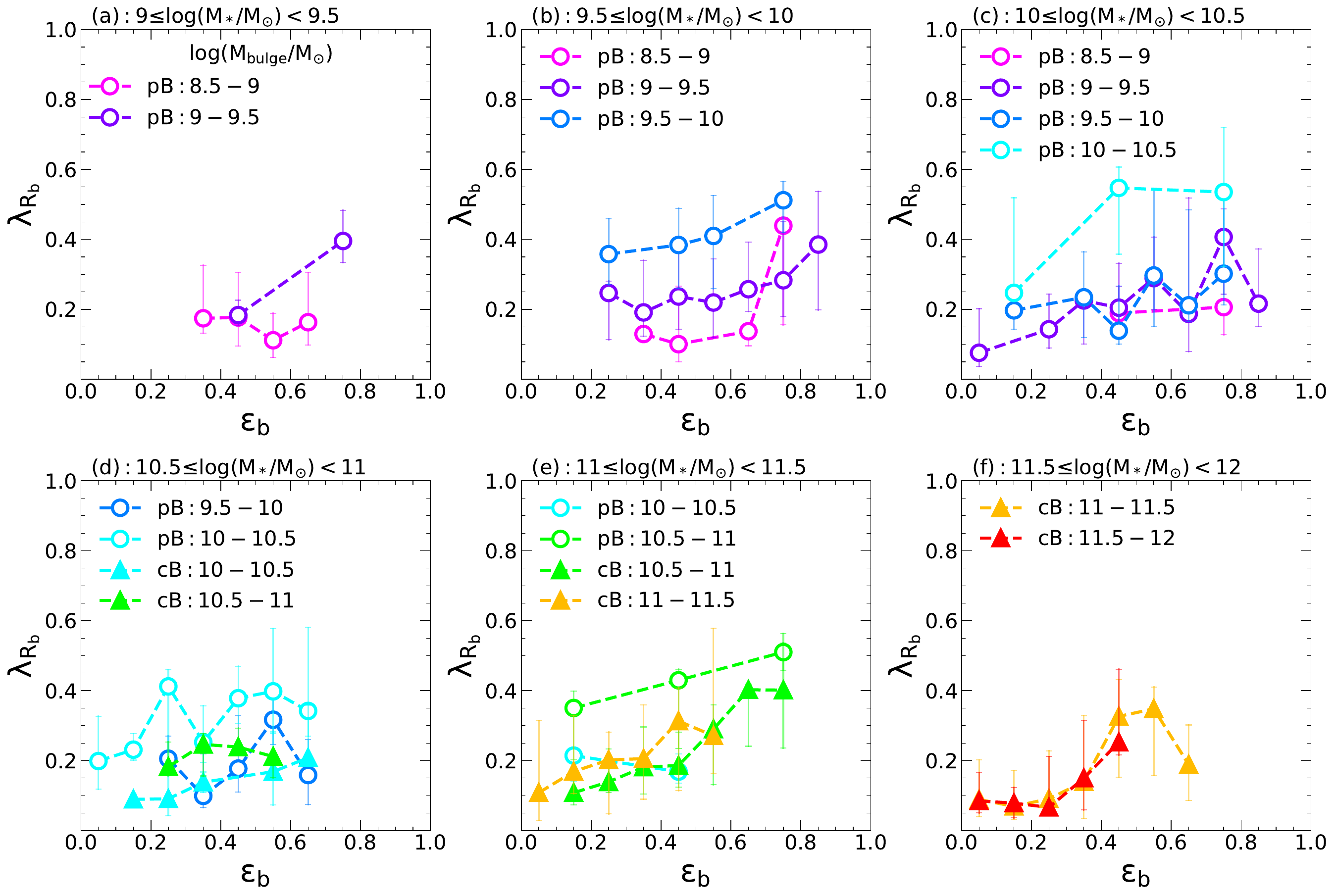}
	\caption{The $\lambda_{R_b}-\epsilon_{b}$ relation for bulges with different bulge mass as indicated in each panel, for pseudo-bulges and classical bulges. Each panel shows results for a given galaxy stellar mass interval as indicated on top of the panel. Symbols present results for the median relation, and galaxies are binned together in some cases to make sure that at least 20 galaxies are included for calculating each symbol. }
    \label{fig: lambda-b in different bulge mass}
\end{figure*}

In Fig.~\ref{fig: lambda-b in different bulge mass}, we further investigate the dependence of $\lambda_{R_b}-\epsilon_b$ relation on bulge mass. 
Similar to the total stellar mass of galaxy as described in \S \ref{sec: Selection of classical and pseudo-bulge galaxies}, bulge mass is calculated as twice the sum of each pixel stellar mass within the ellipse with effective radius $R_b$. 
In Fig.~\ref{fig: lambda-b in different bulge mass}, each panel shows results for a given galaxy total stellar mass interval, and different colored symbols and lines in each panel are the results for subsamples with different bulge masses as indicated.
For pseudo-bulges, at given galaxy stellar mass interval shown in each panel, more massive pseudo-bulges in general rotate more. 
Comparing pseudo-bulges with similar bulge mass but different galaxy total mass in different panels, pseudo-bulges less massive than $10^{9.5}\rm M_\odot$ seem to have consistent results. For
pseudo-bulges with stellar mass in the range $10^{9.5-10.5}\rm M_\odot$, at given bulge mass, pseudo-bulges in more massive galaxies seem to rotate less. 
For classical bulges in each panel, the dependence on bulge mass is weaker than for pseudo-bulges, with still a trend that more massive bulges rotate more. 

For the galaxy stellar mass in the range $10^{10.5-11.5}\rm M_\odot$ where both types of bulges can be hosted as shown in Fig.~\ref{fig: lambda-b in different bulge mass}, at given bulge mass, pseudo-bulges are more rotational dominant than classical bulges. 
Also note that by number fraction in this mass range, most pseudo-bulges have mass range of $10^{9.5-11}\rm M_\odot$, while most classical bulges have mass range of $10^{10-11.5}\rm M_\odot$. This results in the similar rotation between classical and pseudo-bulges with the same total stellar mass as seen in Fig.~\ref{fig: lambda-e in different mass}.

\subsection{Central velocity dispersion}
\label{sec: central velocity dispersion}

Observations of classical and pseudo-bulge galaxies with robust bulge type identification indicate that classical bulges have a large central velocity dispersion, while most pseudo-bulges have a lower central velocity dispersion \citep{2016ASSL..418...41F}. 
The commonly used stellar central velocity dispersion $\sigma_0$ is defined as the mean velocity dispersion of a small region in the central of galaxy. 
For about 100 nearby bulge galaxies, \citet{2016ASSL..418...41F} found that $\sigma_0$ of 68\% classical bulge galaxies and only 9.4\% pseudo-bulge galaxies is larger than 130km/s. 
Therefore a value of $\sigma_0=130$ km/s is often proposed as a criterion to select classical bulges. 
For example, \citet{2020ApJ...899...89S} defined classical bulges with $\sigma_0$ more than 130 km/s and pseudo-bulges with $\sigma_0$ less than 90 km/s. For intermediate $\sigma_0$, there exists no criterion for separating bulge types.

For IFU data as we use in this work, $\sigma_0$ is calculated as the flux-weighted second moments of velocity:
\begin{equation}
    \sigma_0 = \sqrt{\frac{\sum_{i=1}^N F_i(\sigma^2_i+V^2_i)}{\sum_{i=1}^N F_i}}
    \label{eq: sigma}
\end{equation}
where $V_i$ and $\sigma_i$ are the line-of-sight velocity and velocity dispersion on the $i$th pixel. $N$ is the total pixel number within a circle of the central part of a galaxy. 
Different radius of the circle is adopted in previous works, which is $0.25R_e$ in \citet{2020MNRAS.495.4820L}, 1kpc in \citet{2016ASSL..418...41F}, $0.125R_e$ in \citet{2010MNRAS.408...97K} and $0.1R_e$ in \citet{2012ApJ...754...67F}. 
We have checked that for our sample galaxies, $\sigma_0$ calculated within these radii are pretty similar. In this work, we use $0.25R_e$ to include more galaxies, and calculate $\sigma_0$ for all galaxies with $0.25R_e$ greater than their HWHM of PSF of MaNGA (with a mean value of 1.25 $\arcsec$).

\begin{figure}
	\includegraphics[width=\columnwidth]{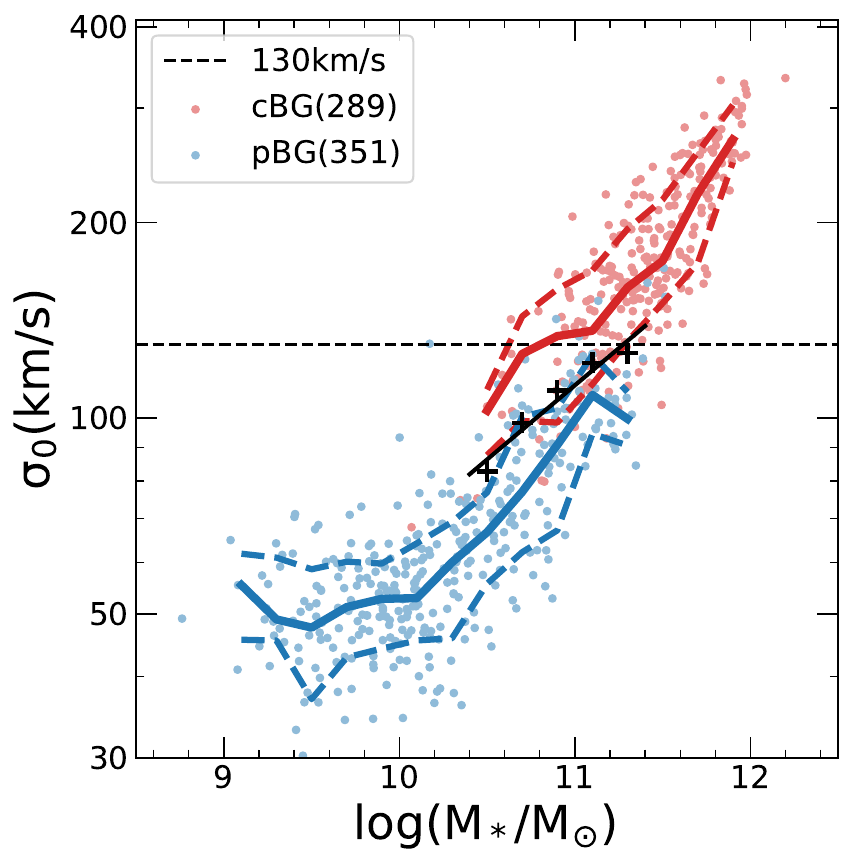}
    \caption{Central velocity dispersion $\sigma_0$ as a function of galaxy stellar mass $M_*$ for classical (red) and pseudo-bulge galaxies (blue). Points are for individual galaxies. Solid lines show the median relations and dashed lines enclose 68\% of the scatter. Black crosses indicate the average of the median $\sigma_0$ for classical and pseudo-bulge galaxies in overlapped stellar mass intervals, which are linearly fitted by the black solid line. The horizontal black dashed line shows  $\sigma_0=130$ $\mathrm{km/s}$, which is commonly used as a criterion to select classical bulges.}
    \label{fig: sigma0-M}
\end{figure}

Fig.~\ref{fig: sigma0-M} gives the results of central velocity dispersion, for galaxies with distinct types of bulges, as a function of galaxy stellar mass. 
For galaxies more massive than $\sim 10^{10}\rm M_\odot$, there exists a strong dependence of $\sigma_0$ on galaxy stellar mass. 
Comparing galaxies with different types of bulges, at given stellar mass, classical bulge galaxies have higher $\sigma_0$, and the two populations are well separated. 
The separation indicates that there exists a criterion of stellar mass dependent $\sigma_0$ that can be roughly used to classify galaxies with classical bulges and with pseudo-bulges. 
In the stellar mass range of $10.4<\log(M_*/\rm M_\odot)<11.4$ where galaxies can host either type of bulge, for each stellar mass interval of 0.2 in logarithmic, we calculate the average of the two median $\sigma_0$ values for classical and pseudo-bulge galaxies. The results are plotted as black crosses in Fig.~\ref{fig: sigma0-M}. These crosses can be fitted by a linear relation as indicated by the black solid line: 
\begin{equation}
    \mathrm{log}(\sigma_0) = 0.23 \times \log(M_*/\rm M_\odot)-0.46
\end{equation}

Compared with the fixed value of $\sigma_0=130$ km/s (the horizontal dashed line in Fig.~\ref{fig: sigma0-M} ) that is previously used to select classical bulges, the equation above provides a more detailed divider that can be used to select both classical and pseudo bulge galaxies. This stellar mass dependent criterion can be applied to galaxies with intermediate central velocity dispersion of [90, 130] km/s, for galaxies with stellar masses of around $10^{10.4-11.4}\rm M_\odot$.

\section{Stellar and cold gas properties}
\label{sec: Stellar population properties}

In this section, we study and compare spatially resolved stellar properties, including age, sSFR and metallicity, for galaxies with classical bulge and pseudo-bulge. 
HI contents of different sample galaxies are also investigated.

\subsection{Properties of stellar population}
\label{Properties of stellar population}

For selected MaNGA galaxies, the spatially resolved stellar population properties including stellar age, metallicity, stellar mass, and flux of H$\alpha$ of each pixel come from the stellar population analysis as described in \S \ref{sec: Selection of classical and pseudo-bulge galaxies}. 
SFR on each pixel is calculated according to H$\alpha$ flux by empirical law of \citet[][Eq.(2)]{1998ARA&A..36..189K}, which represents star formation activities in the last 20 Myr \citep{1998ARA&A..36..189K} in each pixel. 
sSFR is calculated as the ratio of total SFR and total stellar mass within an ellipse.

\begin{figure*}
	\centering
	\includegraphics[width=2.0\columnwidth]{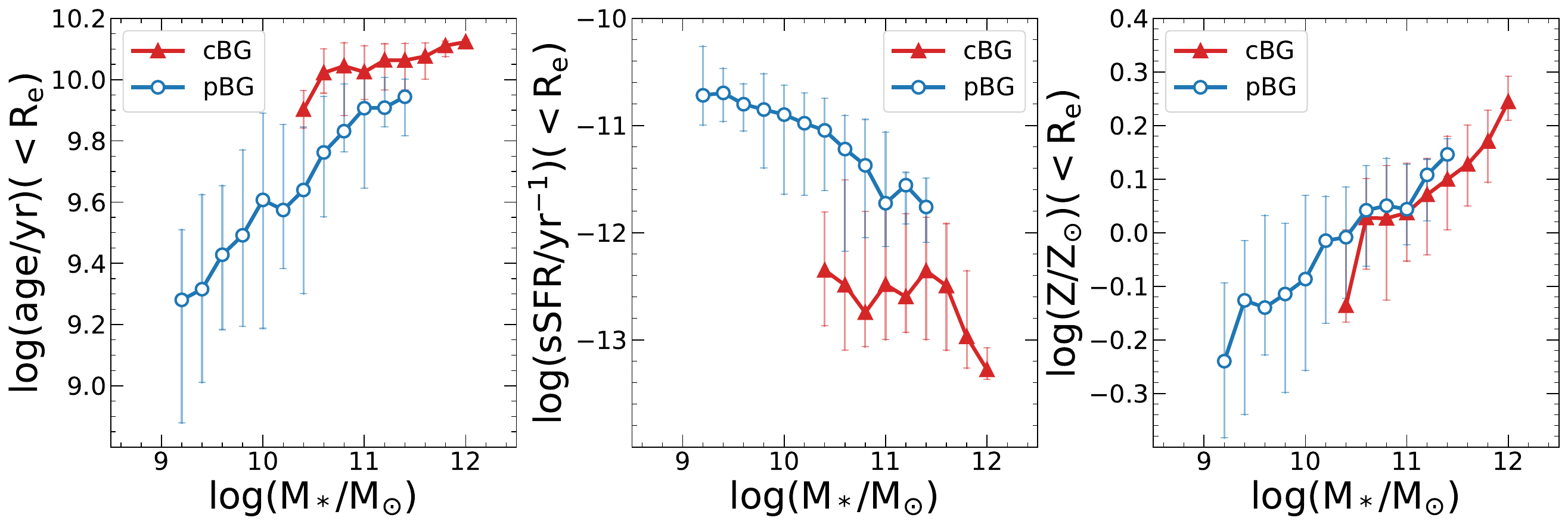}
	\caption{For galaxies with classical bulges (red) and pseudo-bulges (blue), the mass-weighted stellar age (left panel),  sSFR(middle panel), and metallicity (right panel) within $R_e$ as a function of galaxy stellar mass $M_*$. Symbols connected by solid lines present the median relation, and the error bar shows the $1\sigma$ scatter. 
 }
    \label{fig: total population}
\end{figure*}

In Fig.~\ref{fig: total population}, we compare mass-weighted stellar age, specific SFR (sSFR) and stellar metallicity as a function of galaxy stellar mass for the classical and pseudo-bulge galaxies. 
For our galaxy samples, we have checked and find that the average values of the stellar population properties studied within a given radius vary with increasing $R$ and gradually become stable at $R=R_e$ and larger radius. Therefore we use the average values within $R_e$ to represent the global properties of galaxies. 
The left and middle panel of Fig.~\ref{fig: total population} shows that at given stellar mass, classical bulge galaxies have older stellar populations and lower sSFR than pseudo-bulge ones. 
The right panel of Fig.~\ref{fig: total population} indicates that classical and pseudo-bulge galaxies show a similar dependence on total stellar mass, with similar metallicities in most cases.

\begin{figure*}
	\centering
	\includegraphics[width=2.0\columnwidth]{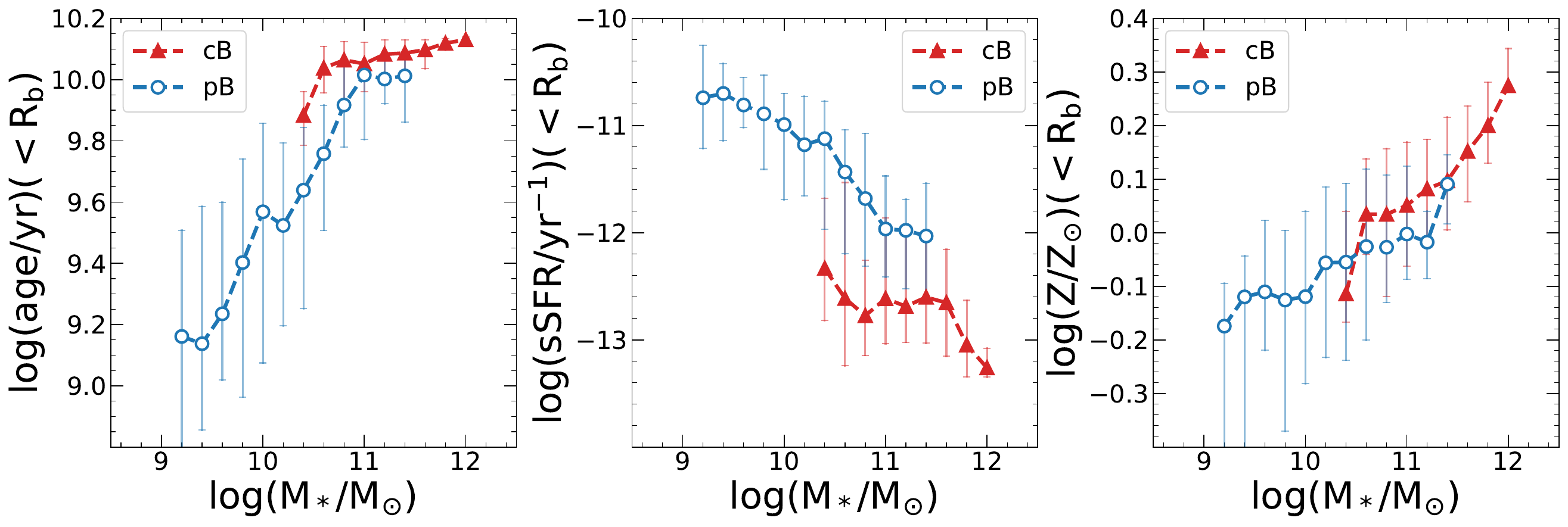}
	\caption{For classical bulges (red) and pseudo-bulges (blue), their average stellar age (left panel), sSFR (middle panel), and metallicity (right panel) as a function of stellar mass of galaxies that host them. }
    \label{fig: bulge population}
\end{figure*}

Fig.~\ref{fig: bulge population} compares the stellar population properties between classical and pseudo-bulge components. 
Classical bulges are in general older and have less efficient star formation than the pseudo-bulges in galaxies with the same stellar mass. 
At stellar mass greater than $10^{11}\rm M_{\odot}$, the difference in stellar age becomes small. 
For stellar metallicity, classical bulges have on average a bit higher metallicity than pseudo-bulges, different from the case for the overall galaxies as indicated in Fig.~\ref{fig: total population}. 

\begin{figure*}
	\centering
	\includegraphics[width=2.0\columnwidth]{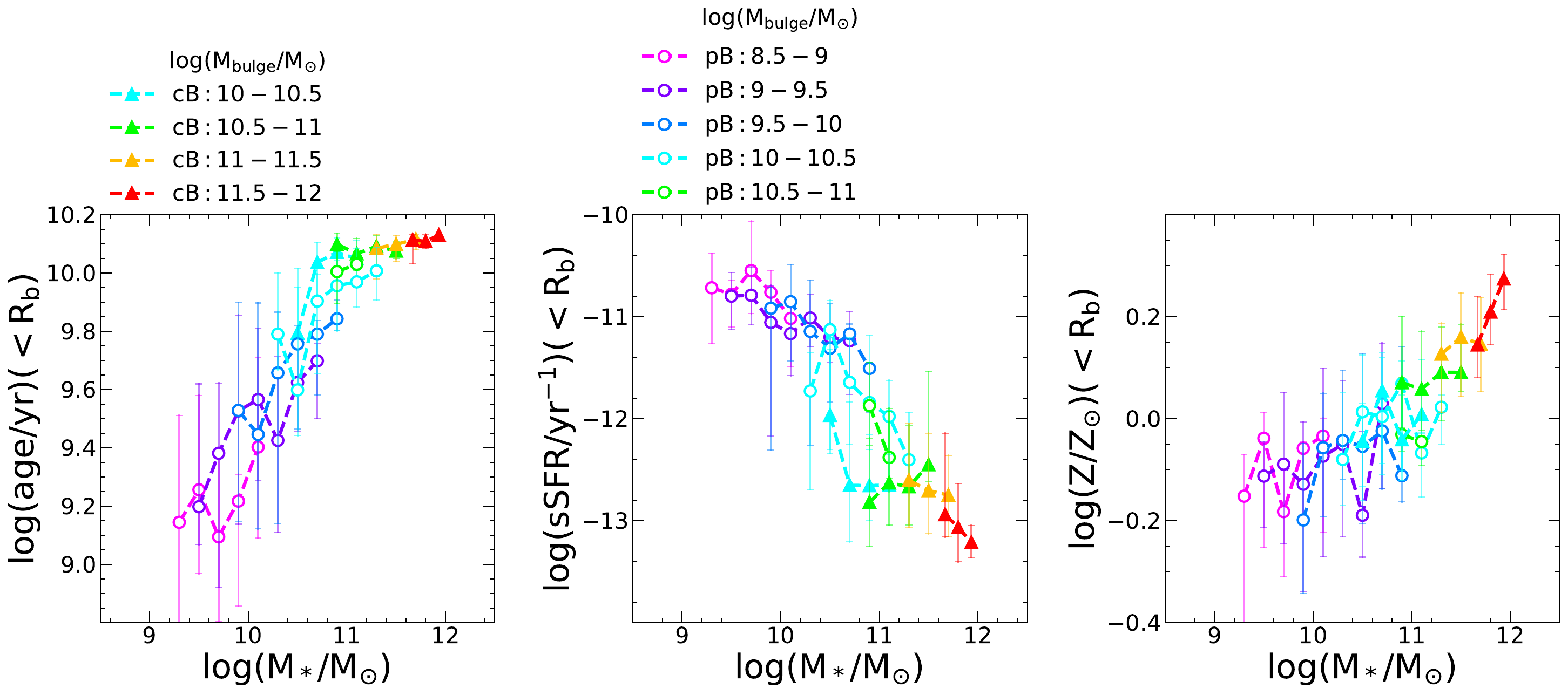}
	\caption{For classical bulges (triangles) and pseudo-bulges (circles), their average stellar age (left panel), sSFR (middle panel), and metallicity (right panel) as a function of stellar mass of galaxies that host them, color-coded by bulge mass as indicated on the top of the panels.  
}
    \label{fig: bulge population vs mb}
\end{figure*}

While strong dependence of stellar properties of bulge component on galaxy stellar mass is seen in Fig.~\ref{fig: bulge population}, we check in Fig.~\ref{fig: bulge population vs mb} whether the dependence is partially due to different bulge mass. 
In Fig.~\ref{fig: bulge population vs mb}, galaxies within a certain stellar mass bin are further divided into subsamples according to their bulge mass, and the results are presented by different colours.
At given stellar mass, the left panel shows that more massive pseudo-bulges have a bit older stellar populations. Apart from that, there exists little dependence on bulge mass for the properties investigated.
Therefore stellar population properties of bulges are most closely related to the stellar mass of their host galaxies rather than their own mass.  

\subsection{Radial profiles of stellar population properties}
\label{profile}

\begin{figure*}
	\centering
	\includegraphics[width=2.0\columnwidth]{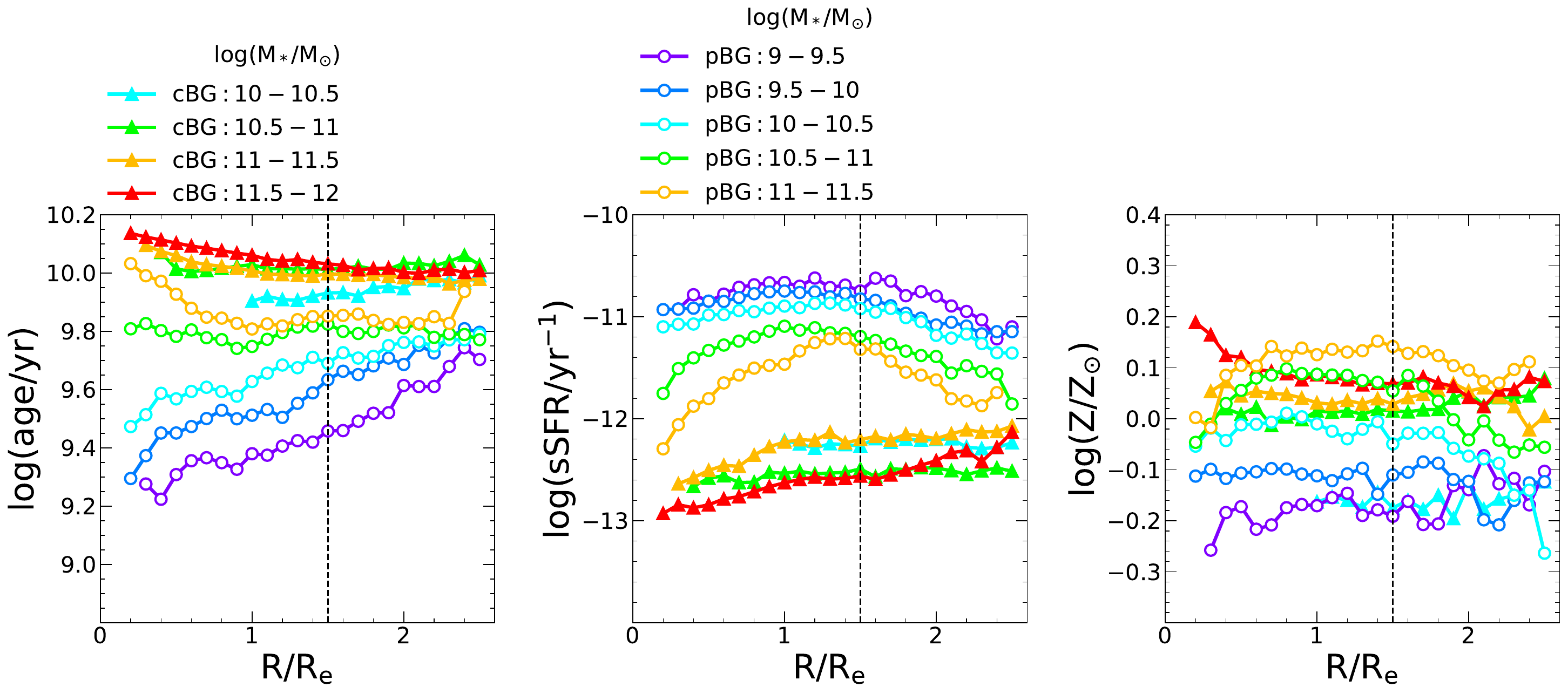}
	\caption{The median radial profiles of stellar age (left panel), sSFR (middle panel), and metallicity (right panel) for classical (triangles) and pseudo-bulge galaxies (circles), in galaxy stellar mass intervals as indicated on top of the panels. Vertical black dotted line marks the radius of $1.5R_e$ in each panel.}
    \label{fig: profile}
\end{figure*}

To investigate in more detail the stellar population in galaxies with classical and pseudo-bulge, especially for the disc components, we compare the profiles of mass-weighted stellar age, metallicity, and sSFR in Fig.~\ref{fig: profile}. 
For each classical and pseudo-bulge galaxy subsample in a certain stellar mass interval, the median stellar population property is calculated and shown at given radius relative to $R_e$. 

In the left panel of Fig.~\ref{fig: profile}, it is evident that classical bulge galaxies have older stellar populations than pseudo-bulge galaxies at all radii, for all stellar mass intervals. 
For classical bulge galaxies, the age profile shows a decreasing trend for massive ones and a flatter trend for lower mass ones. 
For pseudo-bulge galaxies more massive than $10^{10.5}\rm M_{\odot}$ (green and orange circles), the stellar age is older at the center and exhibits a decreasing trend till the radius of around $R_e$. Beyond this radius there exists a slight increase at $1.5R_e$, and the profiles become flat towards outer radii. 
For pseudo-bulge galaxies with stellar mass lower than $10^{10.5}\rm M_{\odot}$, the stellar age continuously increases from center to edge. 
The change of age profile slope with galaxy mass has been found in previous works \citep{2018MNRAS.481.5580F,2020A&A...635A.177B,2022MNRAS.514.6141J,2022MNRAS.514.6120J,2023MNRAS.526.1022L}. 
For example, for face-on late-type galaxies in the stellar mass range $10^{8.9-11.5}\rm M_{\odot}$ in the CALIFA survey, \citet{2020A&A...635A.177B} showed that the inversion of stellar age profile from positive to negative trend occurs at $M_*\sim10^{10}\rm M_{\odot}$. 

In the middle panel of Fig.~\ref{fig: profile}, the behavior of sSFR profiles is generally consistent with the stellar age profiles, with low sSFR corresponding to large stellar age and vise versa. 
For classical bulge galaxies, slopes of sSFR profiles are slightly increasing or stay flat. 
For pseudo-bulge galaxies more massive than $10^{10.5}\rm M_{\odot}$, the sSFR profiles show steep going up and down with increasing radius. 
For pseudo-bulge galaxies in the two lowest stellar mass bins, the sSFR profiles are almost flat. 
Some other studies also focus on the sSFR profile of MaNGA galaxies \citep{2021ApJ...909..120S,2018MNRAS.477.3014B,2022MNRAS.513..389R}.
In particular, \citet{2018MNRAS.477.3014B} found that main sequence galaxies with mass in the range $10^{10-11.5}\rm M_{\odot}$ have an increasing and then decreasing sSFR profile, similar as seen for our (massive) pseudo-bulge galaxies. 
They also studied the sSFR profile of green valley galaxies with stellar mass in the range $10^{10-11.5}\rm M_{\odot}$, which shows an increasing and then flat trend.

The right panel of Fig.~\ref{fig: profile} presents profiles of stellar metallicity. 
For most subsamples, the profile remains flat, especially at intermediate radius, consistent with previous works studying S0 and late-type galaxies \citep{2017MNRAS.465.4572Z,2018MNRAS.481.5580F,2022MNRAS.514.6120J,2022MNRAS.514.6141J,2023MNRAS.526.1022L}.
Exceptions exist for a few massive bins of both bulge type galaxies. Most massive galaxies with classical bulges have obviously higher metallicity in the inner region. Pseudo-bulge galaxies of mass greater than $10^{10.5}\rm M_{\odot}$ have lower metallicity in both the inner and outermost part, which is consistent with the more detailed study of the Milky Way by \citet{2023NatAs...7..951L}, which found a $\Lambda$-shape light-weighted metallicity profile. 
Note that \citet{2023NatAs...7..951L} have checked that Milky Way-mass star-forming galaxies without selecting bulge types in the MaNGA survey in general do not have broken metallicity profiles. This indicates that pseudo-bulge may be critical or closely related to the $\Lambda$-shape metallicity profile of the Milky Way.
Comparing galaxies with two types of bulges, at given stellar mass, pseudo-bulge galaxies have obviously higher metallicity than classical bulge galaxies at intermediate radius, while opposite trend is seen at small radius in the stellar mass range $10^{10.5-11.5}\rm M_{\odot}$. This explains why in this mass range in Fig.~\ref{fig: total population} galaxies with classical bulge have a bit lower metallicity while in Fig.~\ref{fig: bulge population} the opposite is seen.

\subsection{HI gas fraction}
\label{sec: HI gas fraction}

While classical bulge galaxies have older age and lower sSFR than pseudo-bulge ones with the same stellar mass as presented in Fig.~\ref{fig: total population}, we compare their cold gas properties by looking at their HI fraction using HI-MaNGA data. 
HI-MaNGA is an HI (21 cm) follow-up program of MaNGA survey galaxies \citep{2019MNRAS.488.3396M,2021MNRAS.503.1345S}. The latest third data release value-added catalogue contains 6632 sources, where 3274 galaxies of them with $z<0.06$ are provided by ALFALFA catalogue \citep{2018ApJ...861...49H}, and the remaining 3358 galaxies with $z<0.05$ are observed by the Green Bank Telescope.

\begin{figure}
	\includegraphics[width=\columnwidth]{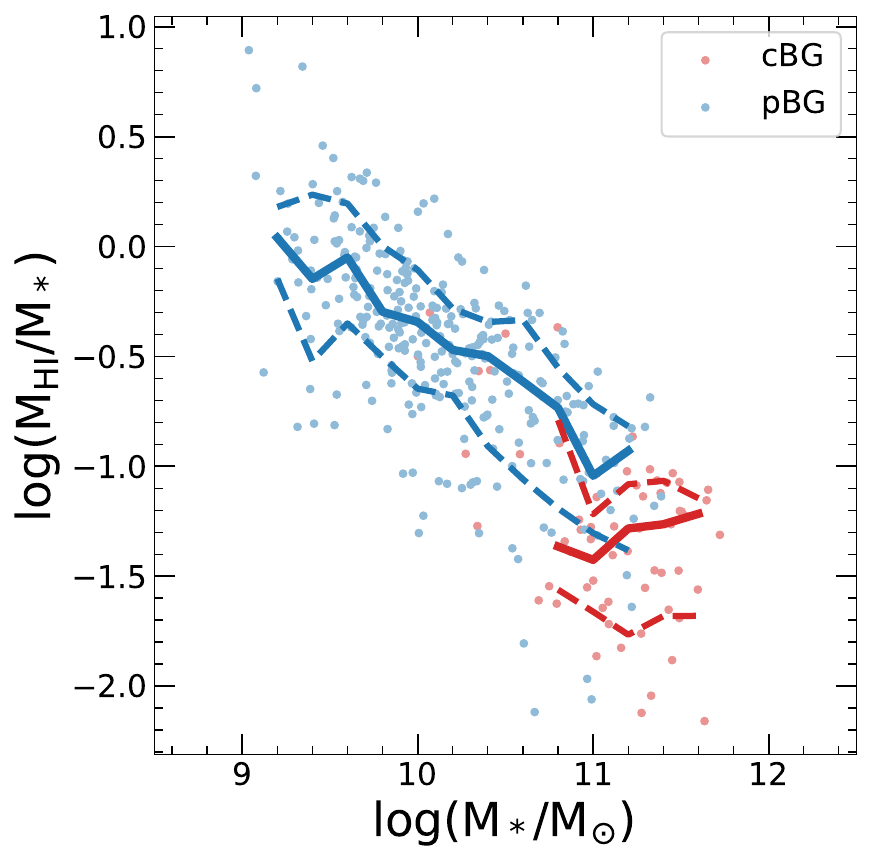}
	\caption{HI fraction as a function of galaxy stellar mass for classical (red) and pseudo (blue) bulge galaxies with HI detection in HI-MaNGA. The scatter points are individual galaxies. The solid lines show the median relation for bins with more than 5 galaxies, and the dashed lines enclose the 68\% distributions. }
    \label{fig: HI mass}
\end{figure}

For our sample galaxies, 176 galaxies with classical bulge are observed by HI-MaNGA, of which 62 (35.23\%) have HI detection. 374 of our pseudo-bulge galaxies are included in the HI-MaNGA catalogue, among which 293 (78.34\%) are HI detected. The HI detection rate of pseudo-bulge galaxies is higher, as expected, since they have more active star formation than classical bulge galaxies. 
In Fig.~\ref{fig: HI mass}, for classical and pseudo-bulge galaxies that have HI detection, we check and compare their HI fraction, the ratio of HI-mass to total stellar mass. The HI gas content of pseudo-bulge galaxies is higher than classical bulge galaxies, in stellar mass range $10^{10.7-11.3}\rm M_{\odot}$.

Fig.~\ref{fig: total population} and Fig.~\ref{fig: bulge population} indicate that in this mass range, classical bulges have a bit older stellar populations and lower sSFR than pseudo-bulges, while the difference for the whole galaxies is obviously larger. 
Also in Fig.~\ref{fig: profile}, larger differences are seen in the disc component than in the inner part of galaxies, for all the stellar properties investigated. 
All these results indicate that discs of pseudo-bulge galaxies are younger, have more active star formation, and are rotating more (Fig.~\ref{fig: lambda of disk different mass}), compared with the discs of classical bulge galaxies, and the difference between discs are larger than that between bulges. \citet{2022A&A...666A.170Q} also found that discs of pseudo-bulge galaxies are bluer than those of classical bulge galaxies with similar stellar mass. Therefore, disc components may contribute more than the bulge components to the higher HI fraction in pseudo-bulge galaxies. In addition, while bulge components tend to be passive and possibly contain little HI gas, the smaller bulge fraction of pseudo-bulge galaxies possibly plays a role in their higher HI context. Future spatially resolved HI observations can provide more information about the direct connection between HI content and different components of galaxies.

\section{Conclusions and discussions}
\label{sec:Conclusions}

In this paper, we compare properties of classical and pseudo-bulges and properties of their hosting galaxies selected from the MaNGA survey. The galaxies with classical bulge and pseudo-bulge are selected based on the S$\mathrm{\acute{e}}$rsic index n of bulge component from the two-component decomposition catalogue provided by \citet{2022MNRAS.509.4024D}, as well as the position of bulges on the Kormandy diagram. 

Our main conclusions are summarized as follows:

- By looking at the $\lambda-\epsilon$ relations of bulges and their hosting galaxies, we find that galaxies hosting pseudo-bulges rotate faster than those with classical bulges at given stellar mass, which is due to a faster-rotating disc, as well as a smaller relative size of bulges for pseudo-bulge galaxies. The $\lambda-\epsilon$ relations for the two types of bulges are similar, and have little dependence on galaxy stellar mass.  

- We propose a new criterion of central velocity dispersion to classify galaxies with classical bulges from those with pseudo-bulges in the mass range of $10.4<\mathrm{log}(M_*/\rm M_\odot)<11.4$, with a divider of: $\mathrm{log}(\sigma_0) = 0.23 \times \mathrm{log}(M_*/\rm M_\odot)-0.46$. Galaxies with larger/smaller central velocity dispersion than the divider can be selected as hosting classical/pseudo-bulges. This mass-dependent separation provides additional criterion apart from the fixed value of $130$ km/s ($90$ km/s) previously used to select classical (pseudo) bulge galaxies  \citep[e.g.][]{2020ApJ...899...89S}. 

- At given galaxy stellar mass, classical bulges and their galaxies have older stellar populations and lower sSFR than pseudo-bulges and their galaxies. Metallicity of classical bulges is higher than pseudo-bulges, but the galaxies hosting classical bulges as a whole have lower metallicity than those hosting pseudo-bulges. All the stellar properties of bulges investigated have strong dependence on galaxy mass, but are almost independent of bulge mass.  

- For the radial profiles of stellar properties, we find that classical bulges have younger age and more active star formation at larger radius, and mostly flat metallicity profiles. Low-mass pseudo-bulge galaxies have increasing stellar age, and roughly flat sSFR and metallicity with increasing radius. Massive pseudo-bulge galaxies have non-monotonic profiles, with youngest stellar age, and largest sSFR and metallicity at radius of around $R_e$ to $1.5R_e$.

Our results indicate that compared with discs of classical bulge galaxies, discs of pseudo-bulge galaxies are younger, have more active star formation, are rotating more, and may contain more HI content (Fig.~\ref{fig: HI mass} and discussions in \S \ref{sec: HI gas fraction}). The difference between discs of galaxies hosting different bulges is larger than that between bulges themselves, both for the stellar population and for the kinematic property (right panel of Fig.~\ref{fig: lambda of disk different mass}). This may give clues to the detailed formation processes of bulges that regulate the growth of central components and the evolution of outer discs. 
On the other hand, as simulations suggest that major mergers could lead to the formation of both elliptical and disc galaxies \citep[e.g.][]{2020MNRAS.493.1375P, 2021MNRAS.507.3301Z}, it is possible that disc-like pseudo-bulges can also form through more violent processes apart from the secular evolution normally considered.

While we use a photometric two-component decomposition catalogue to identify bulge type in this work, methods like multi-component decomposition of galaxy structure \citep{2009MNRAS.393.1531G,2010PASP..122.1397S,2022MNRAS.514.2497K}, full-band spectral decomposition technology \citep{2022MNRAS.514.6120J} and dynamical multi-component decomposition \citep{2024A&A...681A..95J} 
could provide more robust identification of bulge component, but currently only applied to samples with limited numbers of galaxies. When in the future larger samples are available, it may be possible to get more information regarding the correlations that exist between bulge types and their host galaxies/environment \citep{2019MNRAS.484.3865W}, to help us understand more the formation of different types of bulges, and their co-evolution with host galaxies.


\section*{Acknowledgements}
We acknowledge Hu Zou and Zheng Zheng for helpful discussion. This work is supported by the National SKA Program of China (Nos.
2022SKA0110200, 2022SKA0110201), the National Natural Science Foundation of China (NSFC) (grant Nos. 11988101, 11903046), the National Key Program for Science and Technology Research Development of China 2018YFE0202902, the Beijing Municipal Natural Science Foundation (No. 1242032), and K.C.Wong Education Foundation. 
JG acknowledges support from the Youth Innovation 
Promotion Association of the Chinese Academy of 
Sciences (No. 2022056) and the science research grants from the China Manned Space Project.
KZ acknowledges the support from the Shuimu Tsinghua Scholar Program of Tsinghua University.

Funding for the Sloan Digital Sky 
Survey IV has been provided by the 
Alfred P. Sloan Foundation, the U.S. 
Department of Energy Office of 
Science, and the Participating 
Institutions. 
SDSS-IV acknowledges support and 
resources from the Center for High 
Performance Computing  at the 
University of Utah. The SDSS 
website is www.sdss.org.

SDSS-IV is managed by the 
Astrophysical Research Consortium 
for the Participating Institutions 
of the SDSS Collaboration including 
the Brazilian Participation Group, 
the Carnegie Institution for Science, 
Carnegie Mellon University, Center for 
Astrophysics | Harvard \& 
Smithsonian, the Chilean Participation 
Group, the French Participation Group, 
Instituto de Astrof\'isica de 
Canarias, The Johns Hopkins 
University, Kavli Institute for the 
Physics and Mathematics of the 
Universe (IPMU) / University of 
Tokyo, the Korean Participation Group, 
Lawrence Berkeley National Laboratory, 
Leibniz Institut f\"ur Astrophysik 
Potsdam (AIP),  Max-Planck-Institut 
f\"ur Astronomie (MPIA Heidelberg), 
Max-Planck-Institut f\"ur 
Astrophysik (MPA Garching), 
Max-Planck-Institut f\"ur 
Extraterrestrische Physik (MPE), 
National Astronomical Observatories of 
China, New Mexico State University, 
New York University, University of 
Notre Dame, Observat\'ario 
Nacional / MCTI, The Ohio State 
University, Pennsylvania State 
University, Shanghai 
Astronomical Observatory, United 
Kingdom Participation Group, 
Universidad Nacional Aut\'onoma 
de M\'exico, University of Arizona, 
University of Colorado Boulder, 
University of Oxford, University of 
Portsmouth, University of Utah, 
University of Virginia, University 
of Washington, University of 
Wisconsin, Vanderbilt University, 
and Yale University.

\section*{Data Availability}

The Nasa Sloan Atlas catalogue used in this study is available at \url{https://www.sdss.org/dr13/manga/manga-target-selection/nsa/}. The MaNGA kinematics data is publicly available in \url{https://www.sdss4.org/dr17/manga/manga-data/data-access/}. The HI-MaNGA data comes from \url{https://www.sdss.org/dr16/manga/hi-manga/}.



\bibliographystyle{mnras}
\bibliography{references} 






\bsp	
\label{lastpage}
\end{document}